\documentclass[sn-mathphys,Numbered]{sn-jnl}

\usepackage{graphicx}%
\usepackage{multirow}%
\usepackage{amsmath,amssymb,amsfonts}%
\usepackage{amsthm}%
\usepackage{mathrsfs}%
\usepackage[title]{appendix}%
\usepackage{xcolor}%
\usepackage{textcomp}%
\usepackage{manyfoot}%
\usepackage{booktabs}%
\usepackage{algorithm}%
\usepackage{algorithmicx}%
\usepackage{algpseudocode}%
\usepackage{listings}%
\usepackage{url}%

\theoremstyle{thmstyleone}%
\newtheorem{theorem}{Theorem}
\newtheorem{proposition}{Proposition}%

\theoremstyle{definition}%
\newtheorem{example}{Example}%
\newtheorem{remark}{Remark}%

\newtheorem{definition}{Definition}%

\begin{document}

\title[Polynomial GCD certificate for flat bands]
{A polynomial GCD certificate for exact flat bands \\ in finite-range Bloch Hamiltonians}

\author[a,b,c]{\fnm{Ivan} \sur{Damnjanovi\'c}}\email{ivan.damnjanovic@elfak.ni.ac.rs}

\author[d,e]{\fnm{Milan} \sur{Damnjanovi\'c}}\email{yqoq@rcub.bg.ac.rs}

\author[d]{\fnm{Ivanka} \sur{Milo\v sevi\'c}}\email{ivag@rcub.bg.ac.rs}

\author*[f]{\fnm{Dragan} \sur{Stevanovi\'c}}\email{dragan.stevanovic@aasu.edu.kw}

\affil[a]{\orgdiv{Faculty of Electronic Engineering}, \orgname{University of Ni\v s},
           \orgaddress{
           \city{Ni\v s}, 
           \country{Serbia}}}

\affil[b]{\orgdiv{FAMNIT},
          \orgname{University of Primorska}, 
          \orgaddress{
          \city{Koper}, 
          \country{Slovenia}}}

\affil[c]{\orgname{Diffine LLC}, 
          \orgaddress{
          \city{San Diego, CA}, 
          \country{USA}}}

\affil[d]{\orgdiv{Faculty of Physics}, \orgname{University of Belgrade},
          \orgaddress{
          \city{Belgrade}, 
          \country{Serbia}}}

\affil[e]{\orgname{Serbian Academy of Sciences and Arts},
          \orgaddress{
          \city{Belgrade}, 
          \country{Serbia}}}

\affil[f]{
          \orgname{Abdullah Al-Salem University},
          \orgaddress{
          \city{Khaldiya}, 
          \country{Kuwait}}}

\abstract{We formulate a polynomial GCD certificate for exact flat bands in finite-range periodic tight-binding Hamiltonians.  Writing the characteristic polynomial of the Bloch Hamiltonian as a Laurent polynomial
\(
P_L(\mathbf{z},\lambda)=\det(\lambda I-H_B(\mathbf{z}))=\sum_{\mathbf{t}}c_{\mathbf{t}}(\lambda)\mathbf{z}^{\mathbf{t}},
\)
we show that the monic greatest common divisor \(G_L(\lambda)=\gcd_{\mathbf{t}}c_{\mathbf{t}}(\lambda)\) is 
precisely the maximum factor of \(P_L\) that depends only on the energy variable.  
Its roots are exactly the exact flat-band energies, 
and their multiplicities give common algebraic multiplicities of these flat bands throughout the Brillouin zone.
The coefficient-vanishing criterion underlying this statement is known in the flat-band and periodic-graph literature; 
the contribution emphasized here is the compact GCD formulation, its unit cell and Bloch-gauge invariance, 
and its use as a symbolic computation tool for hopping parameter engineering.  
The method is illustrated on kagome, dice and octahedron-chain examples, including weighted kagome and dice lattices.  
The certificate detects exact dispersionless eigenvalues; 
compact localized states, band touching and topological character 
must be analyzed in a subsequent eigenvector or projector calculation.}


\maketitle

\section{Introduction}
\label{sc-intro}

Flat bands are electronic or wave bands whose single-particle energy is independent of crystal momentum.  In the ideal tight-binding limit the kinetic energy within such a band is quenched, so that perturbations, disorder, topology and interactions can dominate the physics.  This makes flat bands useful both as exactly solvable limits and as design targets for correlated and topological phases.  Recent realizations and proposals span quantum materials and moir\'e systems~\cite{NatPhys-20-Zhang,NatPhys-21-Lisi,PRB-24-Liu,PRB-20-Danieli,NatMat-20-Kang,NAT-Balents-20,PRL-Ezzi-24}, topological materials~\cite{NatPhys-22-Calugaru,PRL-20-Ma,PRB-12-Wang}, photonic and other artificial platforms~\cite{Nature-23-Yang,AdvPhys-Leykam-18,Nano-24-Danieli}, and mechanisms for superconductivity and magnetism~\cite{PRL-20-Xie,PRL-20-Lee}.  Reviews of artificial, photonic and singular flat band systems emphasize that exact finite-range tight-binding models remain a central theoretical language, even when experimental bands are only nearly flat~\cite{AdvPhys-Leykam-18,AdvPhys-RhimYang-21,Nano-24-Danieli}.

Several complementary methods for constructing or recognizing flat bands are already available.  The classical examples include Lieb-type bipartite lattices~\cite{PRL-89-Lieb}, Mielke's line graph construction and flat-band ferromagnetism~\cite{JPA-Mielke-L73-91,JPA-Mielke-3311-91}, and compact-localized-state (CLS) descriptions based on destructive interference.  Modern variants include flat band generators from CLS data~\cite{PRB-19-Maimaiti,PRB-21-Maimaiti2D,PRB-21-GrafPiechon}, symmetry and latent-symmetry constructions~\cite{PRB-21-Rontgen}, parameter-dependent construction schemes~\cite{PRB-21-Ogata}, and topological quantum chemistry approaches to crystalline flat bands~\cite{NAT-17-Bradlyn,NatCom-17-Po,NatPhys-22-Calugaru}.  The distinction between nonsingular and singular flat bands, which depends on the behavior of Bloch eigenvectors and on band touching with dispersive bands, is especially important for real space topology and boundary modes~\cite{AdvPhys-RhimYang-21,PRB-21-HwangRhimYang}.

This paper develops a complementary algebraic viewpoint.  For a finite-range Bloch Hamiltonian, the characteristic polynomial is a Laurent polynomial in the Bloch factors and an ordinary polynomial in the energy variable.  A flat band at energy \(\lambda_0\) is equivalent to the vanishing of every Laurent coefficient at \(\lambda_0\).  This coefficient-vanishing criterion is not new: it is closely related to the necessary and sufficient conditions of Toikka and Andreanov~\cite{JPA-Toikka-19}, and a rigorous periodic graph formulation appears in Lemma 2.8 of Sabri and Youssef~\cite{JMP-SabriYoussef-23}.  We therefore do not present the coefficient criterion itself as a new principle.  Instead, we emphasize the following useful packaging and consequences:
\begin{enumerate}
\item the greatest common divisor of all Laurent coefficients extracts the maximum energy-only factor of the characteristic polynomial;
\item the roots of this GCD are the exact flat band energies, with their common algebraic multiplicities;
\item the certificate is invariant under ordinary changes of Bloch gauge and primitive cell basis;
\item treating hopping amplitudes as variables turns the certificate into a symbolic workflow for exact flat band engineering.
\end{enumerate}

The method should be viewed as a detector and design aid for exact dispersionless eigenvalues of finite-range tight-binding models.  It does not, by itself, determine whether a detected band is topological, singular, isolated, or spanned completely by compact localized states.  Those questions require subsequent eigenvector, projector, Wilson loop, symmetry indicator, or quantum geometry analyses.  This separation of tasks is useful: the GCD certificate first identifies exact flat energies and parameter constraints, after which the physical character of the detected band can be analyzed by standard flat band tools.

\section{Relation to previous work and scope}
\label{sc-related}

The algebraic approach used below sits between two established viewpoints.  On one side, the CLS perspective starts from localized eigenstates and constructs hoppings that realize them.  It is physically transparent and directly addresses real space localization, but it may require guessing the relevant CLS pattern.  On the other side, the characteristic polynomial perspective starts from the Bloch Hamiltonian and asks when an energy value is a root of the polynomial for every momentum.  Toikka and Andreanov derived coefficient conditions of this type and used them to generate flat band lattices in several low-dimensional settings~\cite{JPA-Toikka-19}.  Sabri and Youssef proved, in the language of periodic graphs, the fact that the Laurent polynomial vanishing on the torus forces all coefficients to vanish~\cite{JMP-SabriYoussef-23}.

The present formulation adds a compact factorization layer.  Rather than fixing a trial energy \(\lambda_0\) and solving all coefficient equations separately, we compute a single polynomial
\(
G_L(\lambda)=\gcd_{\mathbf{t}}c_{\mathbf{t}}(\lambda).
\)
This polynomial is a certificate: it is nonconstant if and only if the model has an exact flat band, and it gives all exact flat band energies at once.  When some hoppings are symbolic, the same coefficients give polynomial constraints for engineering flat bands at a prescribed energy.  This is a modest but practical shift of emphasis, especially for computer algebra screening of finite-range tight-binding graphs with several tunable hoppings.

The scope is deliberately limited.  The Hamiltonians are finite-range and periodic, so the characteristic polynomial is a Laurent polynomial.  The method detects exact flatness, not approximate narrow bandwidth.  It is spectral: it says where exact flat eigenvalues occur, but it does not classify the corresponding eigenvectors.  In particular, fragile topology, Chern character, singular band crossings, and completeness of compact localized states are invisible to \(G_L\) alone and must be diagnosed separately.

\section{Polynomial GCD certificate from the characteristic polynomial}
\label{sc-charpoly}

Let \(L\) be a translationally invariant, \(d\)-dimensional lattice with \(n\) orbitals or sites per unit cell and finite-range hopping.  We say that \(L\) has a flat band at energy \(\lambda_0\) if \(\lambda_0\) is an eigenvalue of the Bloch Hamiltonian for every wave vector in the Brillouin zone.  Thus the energy eigenvalue, not necessarily the eigenvector, is independent of momentum.

Let \(a_1,\dots,a_d\) be primitive lattice vectors.  Denote by \(A\) the matrix of onsite terms and intracell hoppings, and by \(A_{\mathbf{m}}\) the hopping matrix from a reference cell to the cell translated by \(\sum_jm_ja_j\), where \(\mathbf{m}=(m_1,\dots,m_d)\in\mathbb{Z}^d\).  Hermiticity gives the reverse hopping.  The Bloch Hamiltonian has the finite Laurent form
\begin{equation}
\label{eq-bloch-hamiltonian}
H_B(\mathbf{k}) = A + \sum_{\mathbf{m}\in M} \left(e^{i\sum_j m_jk_j} A_{\mathbf{m}}
                                                 + e^{-i\sum_j m_jk_j} A^{\dagger}_{\mathbf{m}}\right),
\end{equation}
where $\mathbf{k} = (k_1, \ldots, k_d)$ and \(M\) is finite.  Setting \(z_j=e^{ik_j}\) gives
\begin{equation}
\label{eq-bloch-hamiltonian-z}
H_B(\mathbf{z}) = A + \sum_{\mathbf{m}\in M} \left(\prod_j z_j^{m_j} A_{\mathbf{m}}
                                                    + \prod_j z_j^{-m_j} A^{\dagger}_{\mathbf{m}}\right),
\end{equation}
with \(\mathbf{z}=(z_1,\dots,z_d)\in (S^1)^d\).  The characteristic polynomial
\begin{equation}
\label{eq-char-poly}
P_L(\mathbf{z},\lambda)=\det(\lambda I-H_B(\mathbf{z}))
\end{equation}
belongs to the Laurent polynomial ring
\(
\mathbb{C}[z_1^{\pm1},\dots,z_d^{\pm1},\lambda].
\)
Hence it can be written uniquely as
\begin{equation}
\label{eq-rewrittal}
P_L(\mathbf{z},\lambda)=\sum_{\mathbf{t}\in T}c_{\mathbf{t}}(\lambda)\mathbf{z}^{\mathbf{t}},
\qquad
\mathbf{z}^{\mathbf{t}}=\prod_{j=1}^d z_j^{t_j},
\end{equation}
where $\mathbf{t} = (t_1, \ldots, t_d)$, \(T\subset\mathbb{Z}^d\) is finite and each \(c_{\mathbf{t}}(\lambda)\in\mathbb{C}[\lambda] \), $c_{\mathbf{t}} \not\equiv 0$. Multiplying by a sufficiently high monomial \(z_1^N\cdots z_d^N\), if desired, converts this Laurent polynomial into an ordinary polynomial without changing the coefficient polynomials except for a reindexing of exponents.

The following criterion is standard, but we include it to fix notation and to make clear why coefficients rather than sampled momenta are sufficient.

\begin{theorem}[Coefficient criterion]
\label{th-main}
Let \(P_L\) be written as in Eq.~(\ref{eq-rewrittal}).  Then \(L\) has a flat band at \(\lambda_0\) if and only if
\begin{equation}
\label{eq-coeff-vanish}
c_{\mathbf{t}}(\lambda_0)=0
\quad\text{for every }\mathbf{t}\in T.
\end{equation}
\end{theorem}

\begin{proof}
If Eq.~(\ref{eq-coeff-vanish}) holds, then \(P_L(\mathbf{z},\lambda_0)=0\) for all \(\mathbf{z}\in(S^1)^d\), hence \(\lambda_0\) is an eigenvalue at every momentum.  Conversely, suppose \(P_L(\mathbf{z},\lambda_0)=0\) on \((S^1)^d\).  After multiplying by a monomial, we obtain an ordinary polynomial in \(z_1,\dots,z_d\) that vanishes on \((S^1)^d\).  Fixing \(z_2,\dots,z_d\) gives a one-variable polynomial in \(z_1\) with infinitely many roots, hence all its coefficients vanish.  Repeating the same argument for \(z_2\), then \(z_3\), and so on, shows that every Laurent coefficient vanishes at \(\lambda_0\).  This is the usual induction argument for Laurent polynomials on the algebraic torus, and is the point made rigorously for periodic graphs in Ref.~\cite{JMP-SabriYoussef-23}.
\end{proof}

\begin{definition}
\label{def-gcd}
The \emph{polynomial GCD indicator} of \(L\) is the monic polynomial
\begin{equation}
\label{eq-gcd-indicator}
G_L(\lambda)=\gcd\{c_{\mathbf{t}}(\lambda):\mathbf{t}\in T \}\in\mathbb{C}[\lambda].
\end{equation}
A constant indicator means that no exact flat band is present.
\end{definition}

\begin{theorem}[Maximum energy-only factor]
\label{th-consequence}
The polynomial \(G_L(\lambda)\) is the maximum factor of \(P_L(\mathbf{z},\lambda)\) that depends only on \(\lambda\).  Equivalently,
\begin{equation}
\label{eq-factorization}
P_L(\mathbf{z},\lambda)=G_L(\lambda)Q_L(\mathbf{z},\lambda)
\end{equation}
for some Laurent polynomial \(Q_L\), and every \(h(\lambda)\in\mathbb{C}[\lambda]\) that divides \(P_L\) also divides~\(G_L\). Consequently, the roots of \(G_L\) are exactly the flat band energies of \(L\). If $r$ is the multiplicity of $\lambda_0$ as a root of \(G_L\), then \(\lambda_0\) is a root of \(\det(\lambda I-H_B(\mathbf{z}))\) of algebraic multiplicity at least \(r\) for every \(\mathbf{z}\in(S^1)^d\), and \(r\) is the largest such multiplicity visible from the characteristic polynomial.
\end{theorem}

\begin{proof}
Since \(G_L\) divides every coefficient \(c_{\mathbf{t}}\), it divides \(P_L\).  Conversely, if \(h(\lambda)\) divides \(P_L\), then the coefficients of \(P_L\), viewed as a Laurent polynomial in \(\mathbf{z}\), are all divisible by \(h\).  Hence \(h\) divides their greatest common divisor.  This proves the maximum factor statement.  The root statement follows from Theorem~\ref{th-main}.  The multiplicity statement follows by applying the same divisibility argument to powers of \(\lambda-\lambda_0\).
\end{proof}

\begin{proposition}[Bloch gauge and primitive cell invariance]
\label{prop-invariance}
The monic indicator \(G_L\) is invariant under shifts of orbital origins inside the unit cell and under unimodular changes of primitive lattice vectors.  More precisely, replacing \(H_B(\mathbf{z})\) by
\(
D(\mathbf{w})^{-1}H_B(\mathbf{w}^{Q})D(\mathbf{w})
\)
where \(D\) is an invertible diagonal monomial matrix and \(Q\in \mathrm{GL}_d(\mathbb{Z})\), leaves \(G_L\) unchanged.
\end{proposition}

\begin{proof}
The diagonal monomial conjugation is a Bloch gauge change and leaves the characteristic polynomial unchanged.  The substitution \(\mathbf{z}=\mathbf{w}^{Q}\) is an automorphism of the Laurent polynomial ring because \(Q\) is unimodular; it reindexes the monomials but does not change the set of coefficient polynomials.  Their monic GCD is therefore unchanged.
\end{proof}

\begin{remark}
The unimodularity assumption in Proposition~\ref{prop-invariance} ensures that 
the change of variables corresponds to a change of primitive basis of the same Bravais lattice, 
rather than to a passage to a nonprimitive supercell. 
Equivalently, for \(Q\in \mathrm{GL}_d(\mathbb{Z})\), 
the substitution \(\mathbf z=\mathbf w^Q\) is an automorphism of the Laurent polynomial ring
\[
\mathbb{C}[z_1^{\pm1},\ldots,z_d^{\pm1}],
\]
and therefore it only relabels the monomial coefficients in the Bloch characteristic polynomial. 
By contrast, using a nonprimitive supercell, 
corresponding for instance to an integer matrix \(Q\) with \(|\det Q|>1\), folds the band structure. 
The set of flat band energies, i.e., the roots of the indicator, is preserved, 
but the powers of the corresponding factors may change, 
because the same physical flat band may be repeated in the folded band structure. 
For this reason, the primitive cell formulation is the most economical one when multiplicities are discussed.
\end{remark}

\begin{algorithm}[t]
\caption{Polynomial GCD flat band certificate}
\label{alg-gcd}
\begin{algorithmic}[1]
\Require finite-range Bloch Hamiltonian \(H_B(\mathbf{z})\)
\Ensure polynomial indicator \(G_L(\lambda)\)
\State compute \(P_L(\mathbf{z},\lambda)=\det(\lambda I-H_B(\mathbf{z}))\)
\State collect \(P_L\) as a Laurent polynomial in \(\mathbf{z}\): \(P_L=\sum_{\mathbf{t}}c_{\mathbf{t}}(\lambda)\mathbf{z}^{\mathbf{t}}\)
\State return the monic greatest common divisor \(G_L(\lambda)\) of all the \(c_{\mathbf{t}}(\lambda)\)
\end{algorithmic}
\end{algorithm}

Thus \(G_L\) serves as a compact exact certificate for flat bands.  The procedure does not require guessing \(\lambda_0\), nor does it require constructing eigenvectors.  In the examples below we compute the Laurent coefficients explicitly to keep the argument transparent.

\begin{figure}[h!]
\begin{center}
\includegraphics[width=0.4\textwidth]{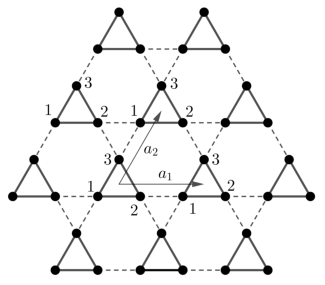}
\qquad
\includegraphics[width=0.4\textwidth]{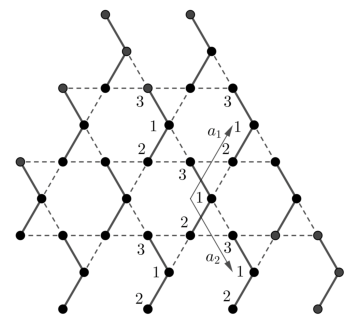}
\end{center}
\caption{Two possible ways of dividing the kagome lattice into unit cells.
Unit cells are shown with bold lines, and intercell hopping with dashed lines.
Variables $a_1$ and $a_2$ denote the primitive lattice vectors.}
\label{fig-kagome}
\end{figure}

\begin{example}
\label{ex-kagome}
The kagome lattice is probably the most well-known lattice with a flat band.
Two possible ways of dividing it into unit cells are shown in Fig.~\ref{fig-kagome}.
The representation on the left leads to the following intra- and intercell hopping matrices:
$$
A=\begin{bmatrix}0&1&1\\1&0&1\\1&1&0\end{bmatrix},
\qquad
A_{(1,0)} = \begin{bmatrix}0&0&0\\1&0&0\\0&0&0\end{bmatrix},
\qquad
A_{(0,1)} = \begin{bmatrix}0&0&0\\0&0&0\\1&0&0\end{bmatrix},
\qquad
A_{(-1,1)} = \begin{bmatrix}0&0&0\\0&0&0\\0&1&0\end{bmatrix}.
$$
After developing,
the characteristic polynomial of its Bloch Hamiltonian can be written in the form
\begin{align*}
\det(\lambda I - H_B(z_1,z_2))
                    = &- (\lambda+2)z_1^1z_2^0 \\
                      &- (\lambda+2)z_1^1z_2^{-1} \\
                      &- (\lambda+2)z_1^0z_2^1 \\
                      &+ (\lambda+2)(\lambda^2-2\lambda-2)z_1^0z_2^0 \\
                      &- (\lambda+2)z_1^0z_2^{-1} \\
                      &- (\lambda+2)z_1^{-1}z_2^1 \\
                      &- (\lambda+2)z_1^{-1}z_2^0,
\end{align*}
so that the gcd of all coefficients $c_{\mathbf{t}}(\lambda)$ is $\lambda+2$.
Hence the kagome lattice has a flat band at $-2$.

The representation on the right of Fig.~\ref{fig-kagome}
leads to different intra- and intercell hopping matrices:
$$
A=\begin{bmatrix}0&1&1\\1&0&0\\1&0&0\end{bmatrix},
\qquad
A_{(1,0)} = \begin{bmatrix}0&1&0\\0&0&0\\0&1&0\end{bmatrix},
\qquad
A_{(0,1)} = \begin{bmatrix}0&0&1\\0&0&1\\0&0&0\end{bmatrix}.
$$
These matrices give a different Laurent representative of the characteristic polynomial, 
but Proposition~\ref{prop-invariance} implies the same polynomial GCD indicator; direct computation again gives \(\lambda+2\).
\end{example}

\begin{figure}[h!]
\begin{center}
\begin{minipage}{0.5\textwidth}
\includegraphics[width=\textwidth]{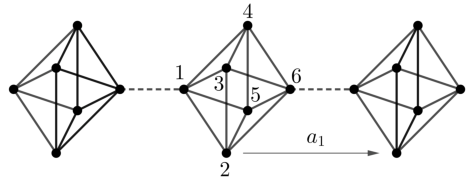}
\end{minipage}
\qquad
\begin{minipage}{0.3\textwidth}
\includegraphics[width=\textwidth]{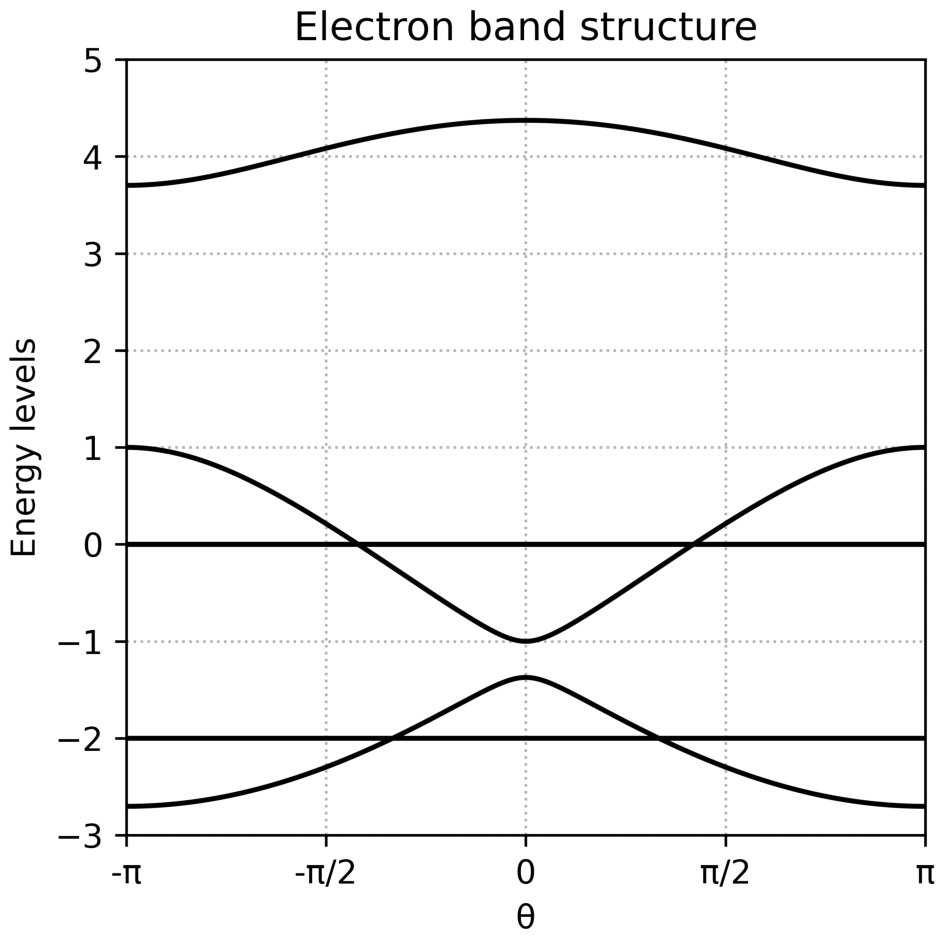}
\end{minipage}
\end{center}
\caption{{\bf Left:} The lattice with an octahedron as the unit cell,
         with intercell hopping represented with the dashed line.
         {\bf Right:} The electron band structure of this lattice.}
\label{fig-octahedron}
\end{figure}

\begin{example}
\label{ex-octahedron}
For the second example we take a somewhat artificial lattice with two flat bands,
shown in the left panel of Fig.~\ref{fig-octahedron}.
It represents the octahedron,
which is itself a double cone over the square,
with two unit cells connected by a hopping between the two neighboring cone sites.
Its unit cell matrix is
$$
A = \begin{bmatrix}
    0&1&1&1&1&0\\
    1&0&1&0&1&1\\
    1&1&0&1&0&1\\
    1&0&1&0&1&1\\
    1&1&0&1&0&1\\
    0&1&1&1&1&0
    \end{bmatrix}
$$
while the intercell matrix $A_1$ has $A_1(6,1)=1$ as its only nonzero entry.
In this case, the characteristic polynomial of the Bloch Hamiltonian
after developing the determinant becomes
\begin{align*}
\det(\lambda I - H_B(z))
  =&-4\lambda^2(\lambda+2)z^1 \\
   &+\lambda^2(\lambda+2)(\lambda^3 - 2\lambda^2 - 9\lambda + 2)z^0 \\
   &-4\lambda^2(\lambda+2)z^{-1}.
\end{align*}
The greatest common divisor of the coefficients $c_{\mathbf{t}}(\lambda)$ above is $\lambda^2(\lambda+2)$,
showing that this lattice has a simple flat band at $-2$ and a double flat band at~0, as illustrated in the right panel of Fig.~\ref{fig-octahedron}.
\end{example}

\begin{figure}[h!]
\begin{center}
\includegraphics[width=0.4\textwidth]{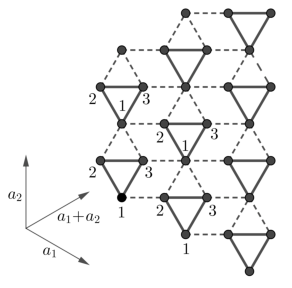}
\end{center}
\caption{The dice lattice with an extra link in the unit cell.
         Intercell hopping is shown with dashed lines.}
\label{fig-dice-twisted}
\end{figure}

\begin{example}
\label{ex-dice-twisted}
The dice lattice is the first reported example of a lattice with a flat band~\cite{PRB-86-Suth}.
To illustrate our method,
in this example we add an extra link to its unit cell between sites 2 and 3 (see Fig.~\ref{fig-dice-twisted})
in order to destroy the flat band.
The intra- and intercell hopping matrices are
$$
A = \begin{bmatrix}
    0&1&1\\1&0&1\\1&1&0
    \end{bmatrix},
\qquad
A_{(1,0)} = \begin{bmatrix}
          0&1&0\\0&0&0\\0&0&0
          \end{bmatrix},
\qquad
A_{(0,1)} = \begin{bmatrix}
          0&0&0\\1&0&0\\1&0&0
          \end{bmatrix},
\qquad
A_{(1,1)} = \begin{bmatrix}
              0&0&0\\0&0&0\\1&0&0
              \end{bmatrix},
$$
which leads to the following form of the characteristic polynomial of its Bloch Hamiltonian:
\begin{align*}
\det(\lambda I - H_B(z_1,z_2))
  &= - 1\cdot z_1^2z_2^1 \\
  &- 2(\lambda+1)z_1^1z_2^1 \\
  &- 2(\lambda+1)z_1^1z_2^0 \\
  &- 2(\lambda+1)z_1^0z_2^1 \\
  &+ (\lambda^3 - 7\lambda - 4)z_1^0z_2^0 \\
  &- 2(\lambda+1)z_1^0z_2^{-1} \\
  &- 2(\lambda+1)z_1^{-1}z_2^0 \\
  &- 2(\lambda+1)z_1^{-1}z_2^{-1} \\
  &- 1\cdot z_1^{-2}z_2^{-1}.
\end{align*}
Since $\gcd(\{1, 2(\lambda+1), \lambda^3-7\lambda-4\})=1$,
this lattice does not have a flat band by Theorem~\ref{th-consequence}.
\end{example}

\section{Flat bands through link weight engineering}
\label{sc-engineering}

An important property of our polynomial flat band indicator is that
it is computed from the characteristic polynomial \(P_L\) in Eq.~(\ref{eq-char-poly}) and its Laurent expansion in Eq.~(\ref{eq-rewrittal}),
which in themselves employ additions and multiplications only.
As a consequence,
if we set some of the link weights in~$L$ to be represented by new unknown variables,
the resulting coefficients $c_{\mathbf{t}}(\lambda)$ will represent
polynomials in both $\lambda$ and the newly introduced variables.
By Theorem~\ref{th-main},
such a parameterized lattice has a flat band at some energy level~$\lambda_0$
iff $c_{\mathbf{t}}(\lambda_0)=0$ for each $\mathbf{t}\in T$,
which actually yields a system of polynomial equations $\{c_{\mathbf{t}}(\lambda_0)=0 : \mathbf{t}\in T\}$ in new variables.
Depending on the choice of the lattice and the parameterized link weights,
the resulting system might be solvable with modern symbolic computation packages, such as SymPy,
especially when the number of variables is relatively small.
The computational method for obtaining the polynomial coefficients $c_{\mathbf{t}}(\lambda)$ is 
implemented in SymPy and openly available in~\cite{GitHub}.
More information on solving systems of polynomial equations
can be found in the book~\cite{sturm}.

Again, let us illustrate this setup on a few examples.

\begin{figure}[h!]
\begin{center}
\includegraphics[width=0.8\textwidth]{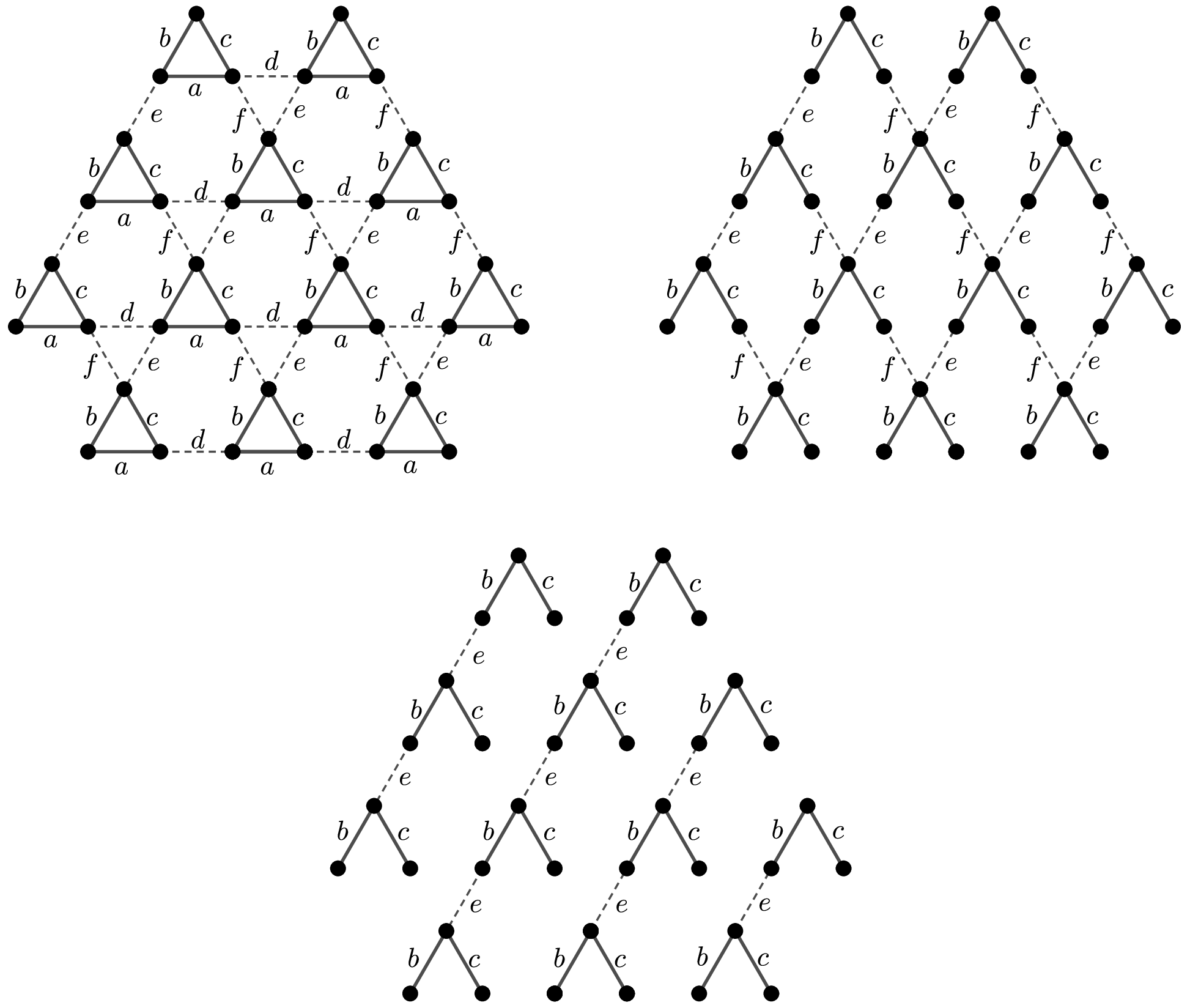}
\end{center}
\caption{{\bf Top left:}
General parameterization of the link weights in the kagome lattice.
The lattice has a flat band for arbitrary $a,b,c$ at any $\lambda<-\max\{\frac{abc}{a^2},\frac{abc}{b^2},\frac{abc}{c^2}\}$,
provided $d,e,f$ are set as in Eqs.~(\ref{eq-d2b})--(\ref{eq-f2b}).
{\bf Top right:}
Setting $a=d=0$ with non-zero $b,c,e,f$ yields a 2D Lieb lattice with a flat band at~0.
{\bf Bottom:}
Setting $a=d=0$ and additionally $f=0$ yields an infinite union of 1D stub lattices, each with a flat band at~0.}
\label{fig-kagome-system}
\end{figure}

\begin{example}
\label{ex-kagome-system}
Let us parameterize link weights in the kagome lattice as shown in Fig.~\ref{fig-kagome-system}.
The intra- and intercell hopping matrices then become
$$
A=\begin{bmatrix}0&a&b\\a&0&c\\b&c&0\end{bmatrix},
\qquad
A_{(1,0)} = \begin{bmatrix}0&0&0\\d&0&0\\0&0&0\end{bmatrix},
\qquad
A_{(0,1)} = \begin{bmatrix}0&0&0\\0&0&0\\e&0&0\end{bmatrix},
\qquad
A_{(-1,1)} = \begin{bmatrix}0&0&0\\0&0&0\\0&f&0\end{bmatrix},
$$
so that the characteristic polynomial of its Bloch Hamiltonian is equal to
\begin{align*}
\det(\lambda I - H_B(z_1,z_2))
= &- (ad\lambda + aef + bcd)z_1^1z_2^0 \\
  &- (cf\lambda + abf + cde)z_1^1z_2^{-1} \\
  &- (be\lambda + ace + bdf)z_1^0z_2^1 \\
  &+ [\lambda^3 - (a^2+b^2+c^2+d^2+e^2+f^2)\lambda - 2abc - 2def]z_1^0z_2^0 \\
  &- (be\lambda + ace + bdf)z_1^0z_2^{-1} \\
  &- (cf\lambda + abf + cde)z_1^{-1}z_2^1 \\
  &- (ad\lambda + aef + bcd)z_1^{-1}z_2^0.
\end{align*}
We want to characterize all parameterizations that lead to a flat band at a certain energy level~$\lambda_0$,
in which case the values of all coefficients in the polynomial above have to be equal to zero.

Suppose first that all link weights $a,b,c,d,e,f$ are non-zero.
From
\begin{align}
\label{eq-coeff1}
ad\lambda_0 + aef + bcd &= 0, \\
\label{eq-coeff2}
cf\lambda_0 + abf + cde &= 0, \\
\label{eq-coeff3}
be\lambda_0 + ace + bdf &= 0,
\end{align}
we have
\begin{align*}
-\lambda_0
 &= \frac{ef}d + \frac{bc}a = \frac{ab}c + \frac{de}f = \frac{ac}b + \frac{df}e \\
 &= \frac{abc}{a^2} + \frac{def}{d^2} = \frac{abc}{c^2} + \frac{def}{f^2} = \frac{abc}{b^2} + \frac{def}{e^2},
\end{align*}
from where
\begin{align}
\label{eq-d2}
\frac{def}{d^2} &= -\lambda_0 - \frac{abc}{a^2}, \\
\label{eq-e2}
\frac{def}{e^2} &= -\lambda_0 - \frac{abc}{b^2}, \\
\label{eq-f2}
\frac{def}{f^2} &= -\lambda_0 - \frac{abc}{c^2}.
\end{align}
Multiplying the last three equalities we obtain
$$
def = - \left(\lambda_0 + \frac{abc}{a^2}\right)
        \left(\lambda_0 + \frac{abc}{b^2}\right)
        \left(\lambda_0 + \frac{abc}{c^2}\right),
$$
so that from Eqs.~(\ref{eq-d2})--(\ref{eq-f2}) we have
\begin{align}
\label{eq-d2b}
d^2 &= \left(\lambda_0 + \frac{abc}{b^2}\right)\left(\lambda_0 + \frac{abc}{c^2}\right), \\
\label{eq-e2b}
e^2 &= \left(\lambda_0 + \frac{abc}{a^2}\right)\left(\lambda_0 + \frac{abc}{c^2}\right), \\
\label{eq-f2b}
f^2 &= \left(\lambda_0 + \frac{abc}{a^2}\right)\left(\lambda_0 + \frac{abc}{b^2}\right).
\end{align}
The free term of $\det(\lambda_0 I - H_B(z_1,z_2))$, i.e., the coefficient of $z_1^0z_2^0$,
then becomes
\begin{align*}
 & \lambda_0^3 - (a^2+b^2+c^2+d^2+e^2+f^2)\lambda_0 - 2abc - 2def \\
={} & \lambda_0^3 - (a^2+b^2+c^2)\lambda_0 \\
-{} & \left[\left(\lambda_0 + \frac{abc}{b^2}\right)\left(\lambda_0 + \frac{abc}{c^2}\right)
        +\left(\lambda_0 + \frac{abc}{a^2}\right)\left(\lambda_0 + \frac{abc}{c^2}\right)
        +\left(\lambda_0 + \frac{abc}{a^2}\right)\left(\lambda_0 + \frac{abc}{b^2}\right)\right]\lambda_0 \\
-{} & 2abc + 2\left(\lambda_0 + \frac{abc}{a^2}\right)
           \left(\lambda_0 + \frac{abc}{b^2}\right)
           \left(\lambda_0 + \frac{abc}{c^2}\right),
\end{align*}
which cancels out completely after simplification,
i.e., the free term also becomes zero with the choice of $d,e,f$ as in Eqs.~(\ref{eq-d2b}--\ref{eq-f2b}).
Note here that $a,b,c$ and $\lambda_0$ can be chosen arbitrarily.
Since we assume that all the link weights are positive,
then the only condition that we obtain from Eqs.~(\ref{eq-d2}--\ref{eq-f2}) is that
\begin{equation}
\label{eq-lambda-bound}
\lambda_0 < - \max\left\{\frac{bc}{a}, \frac{ac}{b}, \frac{ab}{c}\right\}.
\end{equation}
Thus, for any three positive link weights $a,b,c$
the kagome lattice has a flat band at any negative $\lambda_0$ that satisfies~(\ref{eq-lambda-bound}),
provided that the link weights $d,e,f$ are set as given by Eqs.~(\ref{eq-d2b})--(\ref{eq-f2b}).
Setting $a=b=c=1$ and $\lambda_0=-2$,
Eqs.~(\ref{eq-d2b})--(\ref{eq-f2b}) yield $d=e=f=1$ and
the ``usual'' flat band of the kagome lattice when all the link weights are equal to~1.

\bigskip
Assume now that some link weights among $a,b,c,d,e,f$ are equal to zero.
From Eqs.~(\ref{eq-coeff1})--(\ref{eq-coeff3})
we easily obtain the following implications:
\begin{align*}
a=0 &\Rightarrow bcd=0, \\
b=0 &\Rightarrow ace=0, \\
c=0 &\Rightarrow abf=0, \\
d=0 &\Rightarrow aef=0, \\
e=0 &\Rightarrow bdf=0, \\
f=0 &\Rightarrow cde=0.
\end{align*}

Suppose first that there is a pair of links in the same direction
$\{a,d\}$, $\{b,e\}$, $\{c,f\}$
such that both weights from the pair are equal to zero.
Without loss of generality, let us assume that $a=d=0$.
If $\lambda_0\neq0$, then we must also have $be=0$ and $cf=0$ from Eqs.~(\ref{eq-coeff1})--(\ref{eq-coeff3}),
and in either case the lattice fully collapses into disconnected three-site cells.
However, if $\lambda_0=0$,
then both the free term of $\det(\lambda_0 I - H_B(z_1,z_2))$ becomes zero and
Eqs.~(\ref{eq-coeff1})--(\ref{eq-coeff3}) become satisfied for each choice of $b,c,e,f$,
so that the lattice has a flat band at~0.
If all $b,c,e,f$ are non-zero,
the reduced lattice becomes a 2D Lieb lattice~\cite{PRL-89-Lieb}, shown in the middle panel of Fig.~\ref{fig-kagome-system},
while if only one additional value among $b,c,e,f$ is further equal to zero,
the reduced lattice becomes an infinite union of 1D stub lattices~\cite{PRB-17-Rama},
shown in the right panel of Fig.~\ref{fig-kagome-system}.
Note that neither of these two reduced lattices is a line graph.

Let us now assume that we have at most one zero weight in each of the pairs $\{a,d\}$, $\{b,e\}$, $\{c,f\}$,
and without loss of generality, assume that one of $a$ and $d$ is zero.
If $a=0$ and $d\neq 0$, then we must also have $bc=0$,
and once again without loss of generality we can assume that $b=0$ (and $e\neq 0$).
In that case the kagome lattice reduces to an infinite union of the following one-dimensional lattices:
\begin{center}
\includegraphics[width=0.5\textwidth]{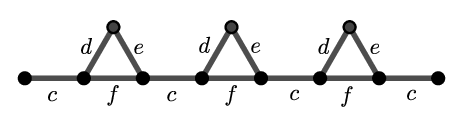}
\end{center}
so that we must have $c\neq 0$ to avoid completely disconnecting the lattice.
Note also that this reduction is still a line graph---this time of the stub lattice.
To obtain a flat band at some $\lambda_0$ in this reduced lattice, we must also have
\begin{align*}
cf\lambda_0 + cde &= 0, \\
\lambda_0^3 - (c^2+d^2+e^2+f^2)\lambda_0 - 2def &= 0.
\end{align*}
From the first equality we have $f \neq 0$ and
$$
\lambda_0 = -\frac{de}f.
$$
Substituting this into the second equality above we obtain
$$
c^2 = \frac{d^2e^2}{f^2} + f^2 - d^2 - e^2 = \left(\frac{d^2}f - f\right)\left(\frac{e^2}f - f\right).
$$
To obtain a positive product here,
we must have that $d$ and~$e$ are either both larger than~$f$ or both smaller than~$f$
(i.e., it is only not allowed that $f$ is between $d$ and~$e$).
Under this condition,
these 1-dimensional lattices have a flat band at $\lambda_0=-\frac{de}f$ for each choice of non-zero $d,e,f$.

Alternatively, if $a\neq 0$ and $d=0$, then we must have $ef=0$,
and this case reduces to the previous one by replacing the roles of $a,b,c$ and $d,e,f$.
\end{example}


\begin{figure}[h!]
\begin{center}
\includegraphics[width=0.45\textwidth]{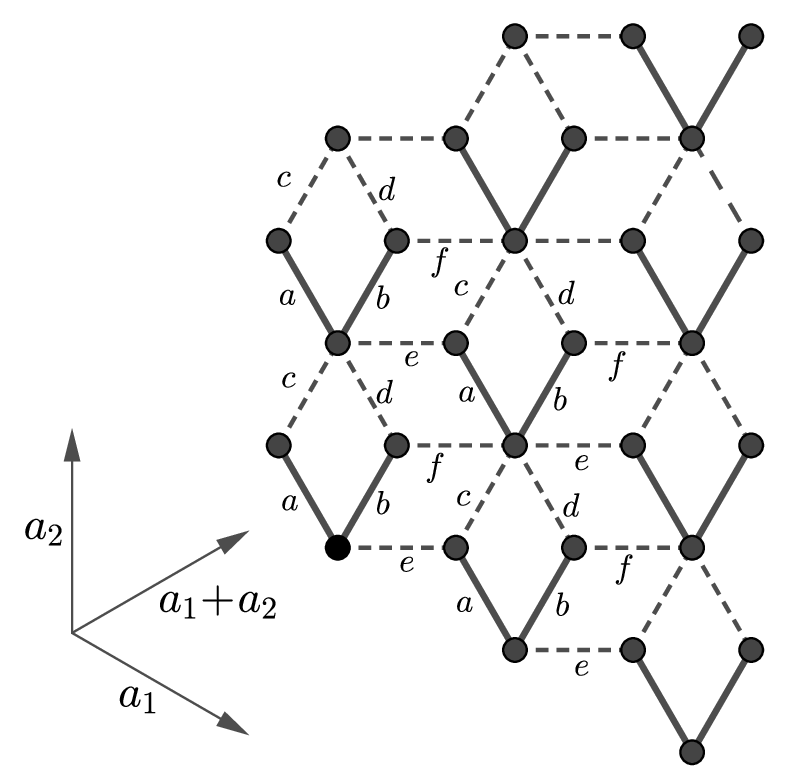}
\end{center}
\caption{General parameterization of link weights in the dice lattice.
         It has a flat band at $\lambda=0$ for each value of the weights $a,b,c,d,e,f$.}
\label{fig-dice-general}
\end{figure}

\begin{example}
\label{ex-dice-system}
In this example we will parameterize the link weights in the dice lattice
in order to characterize the appearance of a flat band.
With the link weights denoted as in Fig.~\ref{fig-dice-general},
we have
$$
A = \begin{bmatrix}
    0&a&b\\a&0&0\\b&0&0
    \end{bmatrix},
\qquad
A_{(0,1)} = \begin{bmatrix}
          0&0&0\\c&0&0\\d&0&0
          \end{bmatrix},
\qquad
A_{(1,0)} = \begin{bmatrix}
          0&e&0\\0&0&0\\0&0&0
          \end{bmatrix},
\qquad
A_{(1,1)} = \begin{bmatrix}
              0&0&0\\0&0&0\\f&0&0
              \end{bmatrix}.
$$
The coefficients of the monomials $z_1^mz_2^n$
in the development of the characteristic polynomial of the Bloch Hamiltonian
now depend also on the variables $a,\dots,f$:
\begin{align*}
\det(\lambda I - H_B(z_1,z_2))
 =&-\lambda(bf + ce) z_1^1z_2^1 \\
  &-\lambda(ae + df) z_1^1z_2^0 \\
  &-\lambda(ac + bd) z_1^0z_2^1 \\
  &+\lambda[\lambda^2\!-\!(a^2\!+\!b^2\!+\!c^2\!+\!d^2\!+\!e^2\!+\!f^2)]z_1^0z_2^0 \\
  &-\lambda(ac + bd) z_1^0z_2^{-1} \\
  &-\lambda(ae + df) z_1^{-1}z_2^0 \\
  &-\lambda(bf + ce) z_1^{-1}z_2^{-1}.
\end{align*}
By Theorem~\ref{th-main}
this lattice has a flat band at some energy level~$\lambda_0$
if all of the above coefficients become zero for $\lambda=\lambda_0$.
Apparently, if $\lambda=0$
then the dice lattice has a flat band at~$0$
for any set of link weights $a,b,c,d,e,f$.

If $\lambda\neq0$,
the system $\{c_{\mathbf{t}}(\lambda)=0 : \mathbf{t}\in T\}$ reduces to
\begin{align}
\label{eq-dice-1}
bf &= -ce, \\
\label{eq-dice-2}
ae &= -df, \\
\label{eq-dice-3}
ac &= -bd, \\
\label{eq-dice-4}
\lambda^2 &= a^2 + b^2 + c^2 + d^2 + e^2 + f^2.
\end{align}
Since we can assume that each link weight is either zero or positive,
Eqs. (\ref{eq-dice-1})--(\ref{eq-dice-3}) imply that actually
$$
ac = ae = ce = bd = bf = df = 0.
$$
Hence at least two values among $\{a,c,e\}$ and at least two values among $\{b,d,f\}$ are equal to zero,
or equivalently,
at most one link from $\{a,c,e\}$ and at most one link from $\{b,d,f\}$ are present in the lattice.
In all such cases,
the starting lattice fully collapses into a disconnected set of $n$-site cells for some $n\in\{1,2,3\}$,
without any intercell links.

Hence a flat band in a general dice lattice exists only at the energy level $\lambda=0$.
\end{example}

\section{After the certificate: eigenstates, touching points and topology}
\label{sc-after}

The GCD certificate is intentionally spectral.  Once a root \(\lambda_0\) of \(G_L\) has been found, the next physical questions concern eigenvectors and projectors.  These can be addressed by studying
\begin{equation}
\label{eq-kernel-after}
K_{\lambda_0}(\mathbf{z})=H_B(\mathbf{z})-\lambda_0 I.
\end{equation}
Over the field of rational functions \(\mathbb{C}(z_1,\dots,z_d)\), the matrix \(K_{\lambda_0}\) has a nontrivial kernel.  A rational kernel vector can be multiplied by a common denominator to give a Laurent-polynomial vector \(v(\mathbf{z})\), and the inverse Fourier transform of such a vector is a compactly supported eigenstate whenever \(v\) is not identically zero.  If every such polynomial Bloch eigenvector vanishes at some points of the Brillouin zone, the flat band is a candidate for singular behavior: compact localized states may fail to span the flat band subspace globally, and noncontractible loop states or boundary modes may be needed~\cite{AdvPhys-RhimYang-21,PRB-21-HwangRhimYang}.  In this sense, the GCD step can be followed by the standard CLS and singular flat band analysis.

Band touching can also be checked algebraically.  After the factorization in Eq.~(\ref{eq-factorization}), the flat band at \(\lambda_0\) touches another band at those momenta for which
\begin{equation}
\label{eq-touching}
Q_L(\mathbf{z},\lambda_0)=0,
\qquad \mathbf{z}\in(S^1)^d,
\end{equation}
possibly with additional rank conditions if several flat factors are present.  In low-dimensional examples, the solutions of Eq.~(\ref{eq-touching}) can be obtained by resultants or by direct symbolic elimination, and then checked on the unit torus.  If the flat band is isolated, one may form its spectral projector and apply the usual Berry phase, Wilson loop, symmetry indicator, or topological quantum chemistry diagnostics~\cite{NAT-17-Bradlyn,NatCom-17-Po,NatPhys-22-Calugaru}.  The polynomial \(G_L\) alone should not be interpreted as a topological invariant.

\section{Computational remarks and limitations}
\label{sc-computation}

The polynomial GCD workflow is exact, but not automatically cheap.  For a unit cell with many sites or many symbolic hoppings, expanding \(\det(\lambda I-H_B)\) can be the computational bottleneck.  Sparse determinant methods, fraction-free elimination, exploitation of lattice symmetries, and collecting terms before full expansion can significantly reduce the cost.  In parameter engineering problems one need not always compute a full symbolic GCD: if the target flat band energy \(\lambda_0\) is prescribed, Theorem~\ref{th-main} gives the polynomial equations \(c_{\mathbf{t}}(\lambda_0)=0\) directly.

The method also has clear physical limitations.  It certifies only exact flat bands of finite-range periodic tight-binding Hamiltonians.  It does not measure approximate bandwidth, stability under perturbations, interaction effects, or disorder sensitivity.  It treats the lattice at the graph and hopping level; embedding geometry, orbital character, spin-orbit coupling, and crystallographic symmetry enter only through the Bloch Hamiltonian supplied as input.  Therefore, for materials applications the GCD certificate should be used as a first algebraic screening step, followed by a physically realistic tight-binding fit or first-principles calculation and by a separate analysis of topology and localization.

\section{Conclusions and outlook}
\label{sc-conclusions}

We have reformulated the characteristic polynomial criterion for exact flat bands as a polynomial GCD certificate.  For a finite-range periodic tight-binding Hamiltonian, the Laurent coefficients of \(\det(\lambda I-H_B(\mathbf{z}))\) have a common factor exactly when the Bloch spectrum contains an exact dispersionless band.  The monic GCD of these coefficients is the maximum energy-only factor of the characteristic polynomial, so its roots give all exact flat band energies and its powers give common algebraic multiplicities throughout the Brillouin zone.  This is a compact way to use the known coefficient-vanishing criterion and is well suited to symbolic computation.

The examples show two uses of the certificate.  First, for a fixed lattice such as the kagome, dice, or the octahedron chain, the GCD immediately detects the exact flat energies and is independent of ordinary choices of Bloch gauge and primitive cell.  Second, when hopping amplitudes are treated as variables, the coefficient equations become design equations for flat band engineering.  The weighted kagome and dice calculations illustrate how the same algebraic object can recover known flat band limits and produce parameter constraints for more general weighted graphs.

Several extensions appear natural.  On the mathematical side, a combinatorial description of the coefficient polynomials \(c_{\mathbf{t}}(\lambda)\) could avoid full determinant expansion for large sparse graphs.  On the computational side, benchmark implementations should compare direct symbolic determinants, sparse elimination, resultants for band touching, and numerical verification on sampled Brillouin zone meshes.  On the physical side, every flat band detected by \(G_L\) should be followed by eigenvector and projector diagnostics: construction of compact localized states, identification of singular points and band touchings, and, where relevant, symmetry indicator or Wilson loop analysis.  With these additions, the polynomial GCD certificate can serve as a useful front end for exact flat band search and hopping parameter engineering in artificial lattices and model tight-binding materials.

\end{document}